\documentclass{aa}  

\usepackage[breaklinks=true,colorlinks,citecolor=blue]{hyperref}
\usepackage{graphicx,animate}
\usepackage{enumerate}
\usepackage{epsfig}
\usepackage{color}
\usepackage{txfonts}
\usepackage{natbib}
\usepackage{multirow}
\usepackage{ulem} 
\usepackage{amssymb}

\usepackage{movie15}

\usepackage{xcolor}

\definecolor{ao}{rgb}{0.0, 0.5, 0.0}

\newcommand{\kmps}{\rm km~s\ensuremath{^{-1} }\,}
\newcommand{\kmskpc}{km~s\ensuremath{^{-1}}~kpc\ensuremath{^{-1} }\,}
\newcommand{\Msun}{M\ensuremath{_\odot}}

\newcommand{\Msunpc}{M\ensuremath{_\odot}~pc\ensuremath{^{-2} }\,}

\newcommand{\Oo}{\displaystyle}

\newcommand{\Rgal}{\ensuremath{\rm R_{gal}}\,}

\newcommand{\hii}{\ensuremath{\rm HII}\,}

\begin{document}

\title{ISM metallicity variations across spiral arms in disk galaxies: \\ the impact of local enrichment and gas migration in the presence of radial metallicity gradient}

\titlerunning{Gas metallicity variations across spiral arms}

\author{Sergey Khoperskov$^{1,2}$, Evgenia Sivkova$^{3}$,  Anna Saburova$^{3,4}$, \\ Evgenii Vasiliev$^{3}$, Boris Shustov$^{3}$, Ivan Minchev$^{1}$, C. Jakob Walcher$^{1}$}

\authorrunning{Khoperskov et al}

\institute{ $^{1}$Leibniz Institut f\"{u}r Astrophysik Potsdam (AIP), An der Sternwarte 16, D-14482, Potsdam, Germany\\ $^{2}$GEPI, Observatoire de Paris, PSL Universit{\'e}, CNRS,  5 Place Jules Janssen, 92190 Meudon, France \\ $^3$Institute of Astronomy, Russian Academy of Sciences, 48 Pyatnitskya St., Moscow, 119017, Russia  \\ $^4$Sternberg Astronomical Institute, Moscow M.V. Lomonosov State University, Universitetskij pr., 13, Moscow 119234, Russia}

\abstract{
Chemical abundance variations in the ISM provide important information about the galactic evolution, star-formation and enrichment histories. Recent observations of disk galaxies  suggest that if {\it large-scale} azimuthal metallicity variations appear in the ISM, they are linked to the spiral arms. In this work, using a set of chemodynamical simulations of the Milky Way-like spiral galaxies, we quantify the impact of gas radial motions~(migration) in the presence of a pre-existing radial metallicity gradient and the local ISM enrichment on both global and local variations of the mean ISM metallicity in the vicinity of the spiral arms. 

In all the models, we find the scatter of the gas metallicity of $\approx0.04-0.06$~dex at a given galactocentric distance. On large scales, we observe the presence of spiral-like metallicity patterns in the ISM which are more prominent in models with the radial metallicity gradient. However, in our simulations, the morphology of the large-scale ISM metallicity distributions significantly differs from the spiral arms structure in stellar/gas components resulting in both positive and negative residual~(after subtraction of the radial gradient) metallicity trends along spiral arms. We discuss the correlations of the residual ISM metallicity values with the star formation rate, gas kinematics and offset to the spiral arms, concluding that the presence of a radial metallicity gradient is essential for the azimuthal variations of metallicity. At the same time, the local enrichment alone is unlikely to drive systematic variations of the metallicity across the spirals.
}


\maketitle

\section{Introduction}\label{sec::intro}
Chemical abundance variations play a key role in our understanding of galactic formation and evolution. In particular, radial abundance gradients in stellar populations are believed to be the result of the inside-out galaxy formation scenario~\citep{1989MNRAS.239..885M,1997ApJ...477..765C,2012MNRAS.426..690B,2013ApJ...773...43B,2018MNRAS.474.3629G}. However, theoretical models also suggest that the radial abundance gradients are affected by various dynamical processes where stars can migrate far away from their birth radii~\citep[see, e.g.,][]{2002MNRAS.336..785S,2008ApJ...684L..79R,2009MNRAS.397.1599Q,2010ApJ...722..112M,2011A&A...527A.147M}. For instance, stellar radial migration caused by the spiral arms and/or spiral/bar resonances overlap results in flattening of the radial gradient~\citep{2009MNRAS.398..591S,2012A&A...540A..56P,2013A&A...554A..47G,2013A&A...558A...9M,2014A&A...572A..92M,2018MNRAS.481.1645M}.

\begin{figure*}
\begin{center}
\includegraphics[width=0.49\hsize]{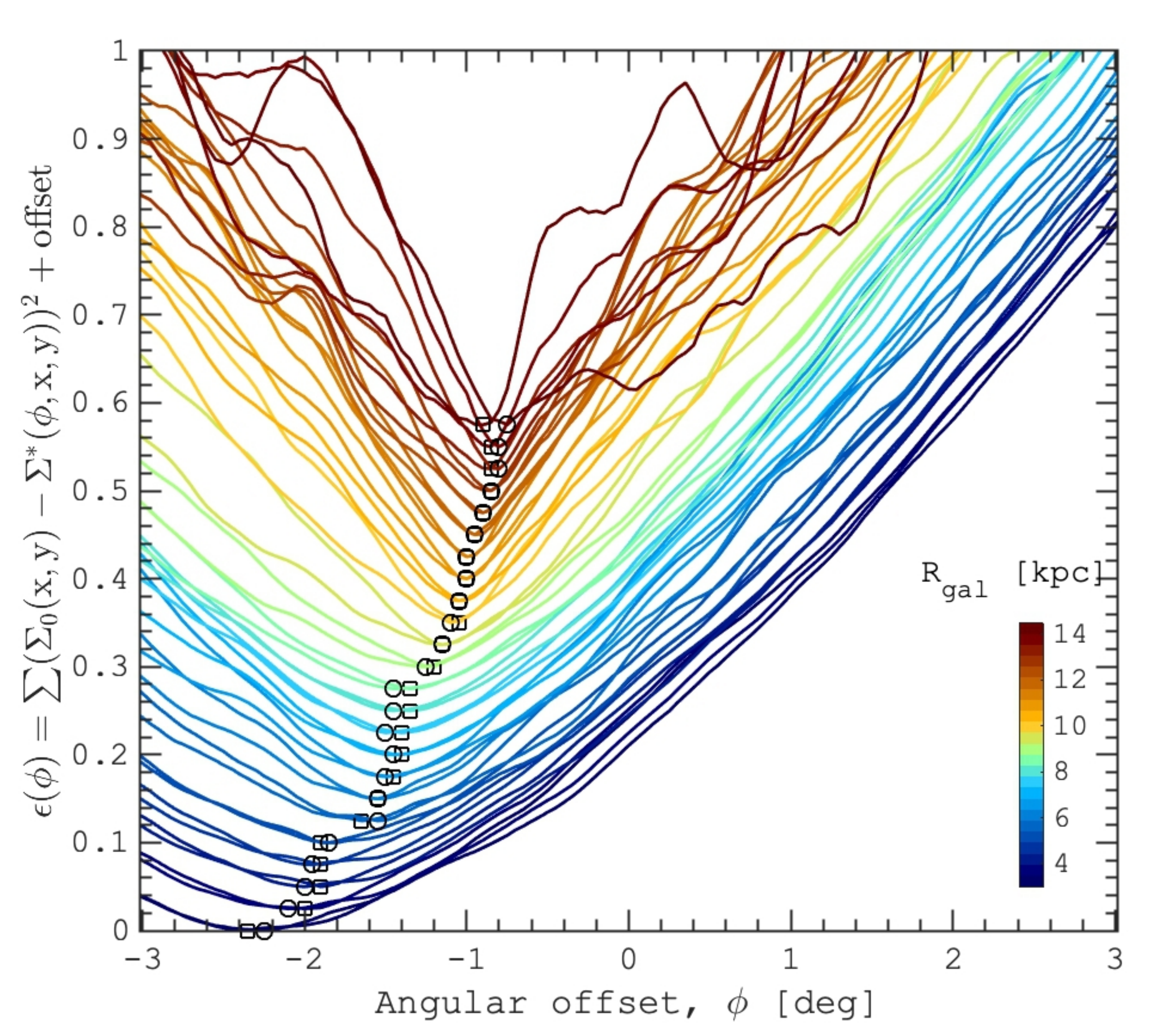}\includegraphics[width=0.49\hsize]{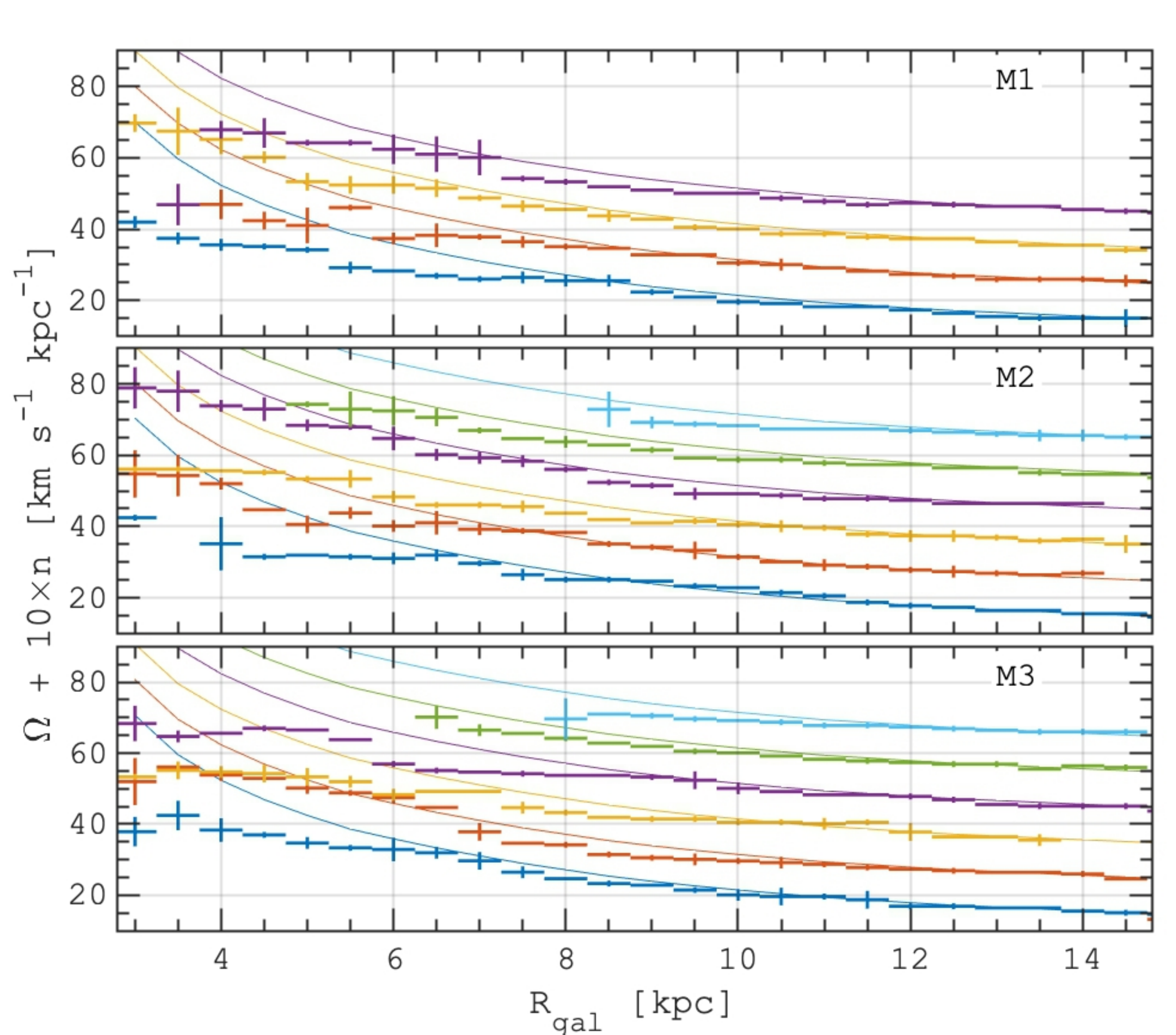}
\caption{{\it Left:} example of the pattern speed measurement for a single spiral arm in model M1. The coloured lines correspond to the sum of the squared-root differences between the stellar density at $t_0=0.6$~Gyr and the ones from $t_0\pm1$~Myr rotated by a certain angle $\phi$~(see Eq.\ref{eq:eps}). The minimum value corresponds to the angle of rotation which being divided by $1$~Myr results in the pattern speed value~(black symbols) at a given \Rgal. {\it Right:} the pattern speed of the spiral arms in different models. Values for the individual spiral arms are shown by the errorbars of different colours. For a better representation, the pattern speed for each arm is shifted vertically by 10~\kmskpc and compared to the circular frequency~(solid lines) which is also shifted by the same values.}\label{fig::Omega0}\label{fig::xi}
\end{center}
\end{figure*}

\begin{figure*}[t]
\begin{center}
\includegraphics[width=1\hsize]{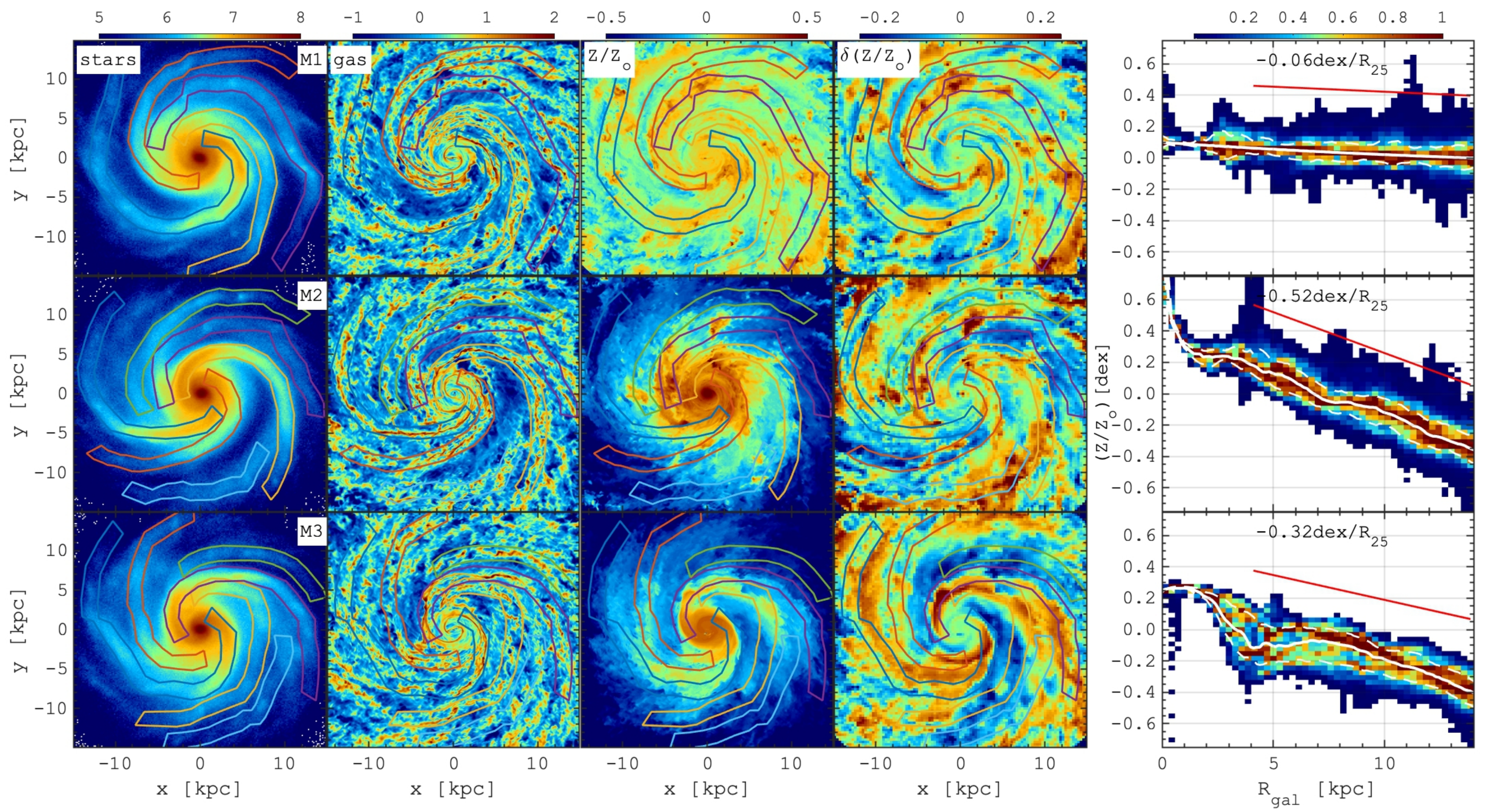}
\caption{Structure of different models after $0.6$~Gyr: M1~(no initial gradient), M2~(both ISM enrichment and gradient) and M3~(only gradient). From left to right: stellar density, gas density, the mean ISM metallicity, residual ISM metallicity~(after subtraction of the radial gradient) and the radial metallicity profile. Different models are shown in different rows. Coloured lines highlight the location of the stellar spiral arms measured as a positive overdensity in the first row~(see Section.~\ref{sec::results} for details). In the right column, the mean radial metallicity trend is shown by the white solid line. Exponential fit of the radial gradient is shown by the red lines~(shifted vertically for better visibility) where $R_{25}=12$~kpc.}\label{fig::fig1}
\end{center}
\end{figure*}

Some models also suggest that spiral arms can cause variations of the mean metallicity of stars in the azimuthal direction~\citep[see, e.g.,][]{2016MNRAS.460L..94G,2018A&A...611L...2K} where the metallicity variations are caused by re-shaping of pre-existing stellar populations with different chemo-kinematical properties in the vicinity of the spiral arms. Alternatively, \cite{2019A&A...628A..38S} developed a new 2D analytic model where stars in the Milky Way~(MW)-type disk inherit systematic~($\approx 0.1$~dex) abundance variations from the ISM, where the latter ones are the result of the enrichment near the arms with the most significant azimuthal variations appearing near the corotation radius. 

Since stars inherit the abundances from the ISM, analysis of the chemical abundance variations of gas plays a fundamental role in understanding the present-day stellar abundance patterns. Observations suggest that the ISM in disk galaxies is well mixed ~(scatter in azimuth of $\approx 0.05$~dex~\citep{Zinchenko2016}) and environmental variations of abundances depend on the observational techniques, disk coverage, and filling factor of individual \hii regions. For instance, analysis of the M101 galaxy has shown evidence for azimuthal variations in gas-phase metallicity~\citep{1996ApJ...456..504K} but later on \cite{2013ApJ...766...17L} found no difference between arm and inter-arm metallicities. Several more recent integral field unit~(IFU) observations are in favour of small but systematic variations of metallicity that appears to correlate with the location of the spiral arms~\citep[see, e.g.,][]{2012A&A...545A..43C,2016ApJ...830....4C,2017A&A...601A..61V, 2018MNRAS.474.1657S}. In particular, by studying the nearby spiral galaxy NGC~1365, \cite{2017ApJ...846...39H} found systematic~($0.2$~dex) azimuthal variations of the \hii region oxygen abundance near the spiral arms imprinted on a negative radial gradient. \cite{2016ApJ...830L..40S} showed that the \hii region oxygen abundances are higher at the trailing edges and lower at the leading edges of the spiral arms which is likely caused by radially outward~(inward) streaming motion at the trailing~(leading) edges of the spiral arms. By analyzing the Very Large Telescope/Multi Unit Spectroscopic Explorer~(VLT–MUSE) data for eight nearby galaxies, \cite{2019ApJ...887...80K} found a low $0.03-0.05$~dex azimuthal abundance scatter where half of the galaxies reveal azimuthal variations which in many cases, however, can not be clearly associated with the spirals. On the other hand, \cite{2016ApJ...827..103K} found that the inter-arm regions of NGC~628 have oxygen abundances similar to the arms which, however, could be the result of poor coverage of the galactic disk. 

A recent study of $45$ nearby spiral galaxies by \cite{2020MNRAS.492.4149S} suggests that $45-65\%$ of galaxies have more metal-rich \hii regions in spiral arms with respect to the inter-arm area. However, in some cases~($5-20\%$, depending on the calibrator), the opposite trend is seen, particularly more metal-poor \hii regions in the spiral arms compared to the inter-arm region. According to \cite{2020MNRAS.492.4149S} more metal rich spiral arms than the inter-arm area are observed in more massive galaxies with grand-design spiral arms.  Finally, \cite{2021arXiv211010697W} have mapped the two-dimensional variations of metals across the disks of 19 nearby galaxies observed with the VLT–MUSE and found no evidence that spiral arms are enriched compared to the disk.

Existing models of the ISM mixing in disk galaxies also have not reached a consensus about the abundance variations in azimuth where both turbulence and gravitational instability act towards homogeneity of the ISM~\citep{2002ApJ...581.1047D, 2003PhRvE..67d6311K, 2018MNRAS.475.2236K, 2012ApJ...758...48Y}, while the large-scale models predict variations driven by spirals~\citep{2016ApJ...830L..40S}\footnote{Note that \cite{2016MNRAS.460L..94G} and \cite{2013A&A...553A.102D} discuss the abundance variations in stellar populations but not in the ISM.}. Therefore, the origin of the azimuthal ISM metallicity variations remains unclear: whether they are real and, if so, are they driven by local self-enrichment~\citep{2017ApJ...846...39H} or by radial flows in the disk~\citep{2016ApJ...830L..40S}? 

In this work, using a set of high-resolution $N$-body hydrodynamical simulations we explore the origin of the ISM abundance variations across the spiral arms of the MW-type disk galaxies. We quantify the impact of the local ISM enrichment by ongoing star formation~(SF) and the transformation of the radial abundance gradient into the azimuthal one due to gas radial displacement~(migration) caused by the spiral arms. The paper is structured as follows. In Sec.~\ref{sec::models} we describe our models setup, subgrid physics and spiral arms kinematics analysis. In Sec.~\ref{sec::results} we discuss both small- and large-scale metallicity variations in the azimuthal direction and along the spirals linking the observed patterns to the properties of individual arms. In Sec.~\ref{sec::conclusions} we discuss and summarize our findings.

\begin{figure}
\begin{center}
\includegraphics[width=1\hsize]{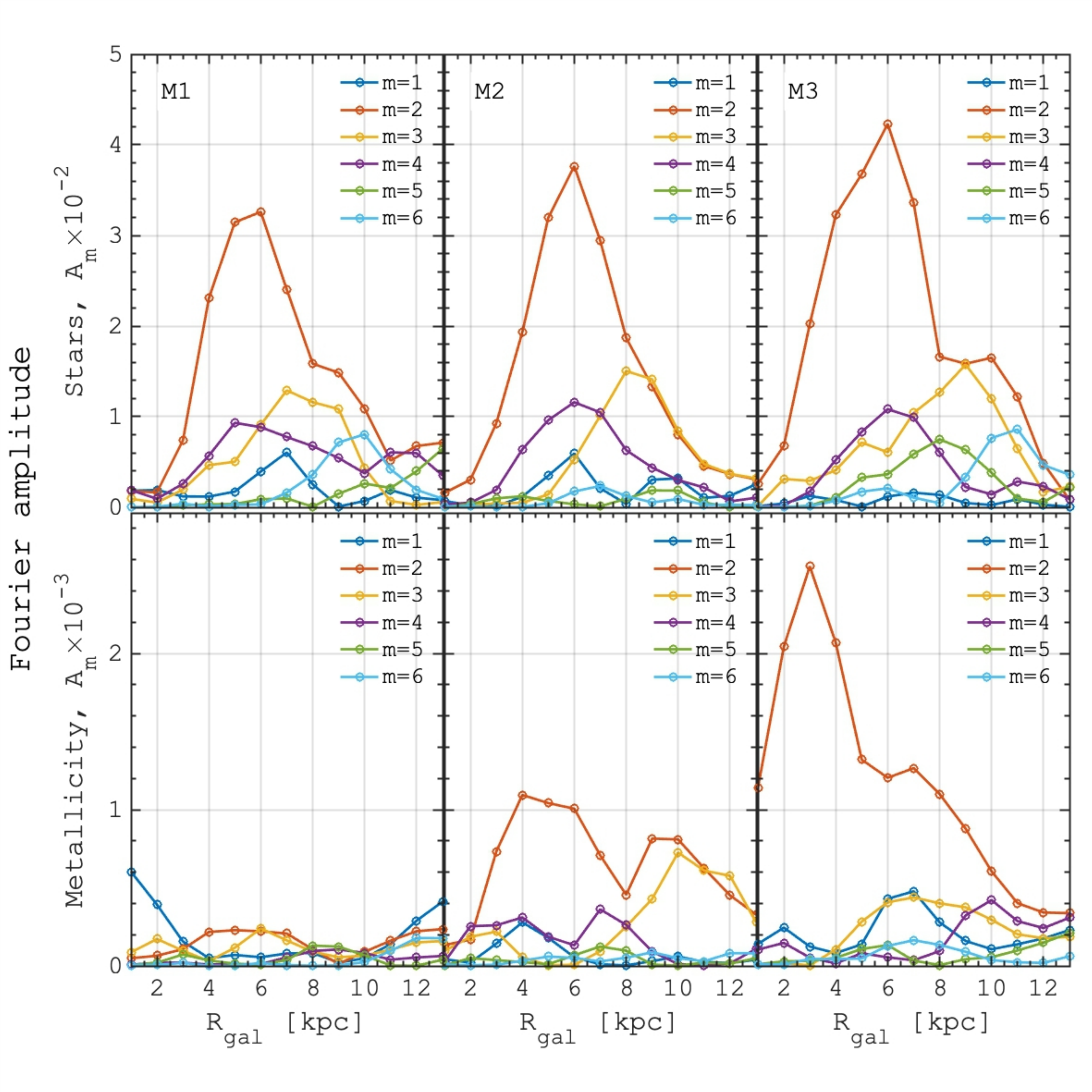}
\caption{Radial structure of the main~($m=1,...,6$) 2D Fourier harmonics computed from the stellar surface density maps~(top) and from the mean metallicity distribution~(bottom).}\label{fig::fourier_amps}
\end{center}
\end{figure}

\begin{figure}
\begin{center}
\includegraphics[width=1\hsize]{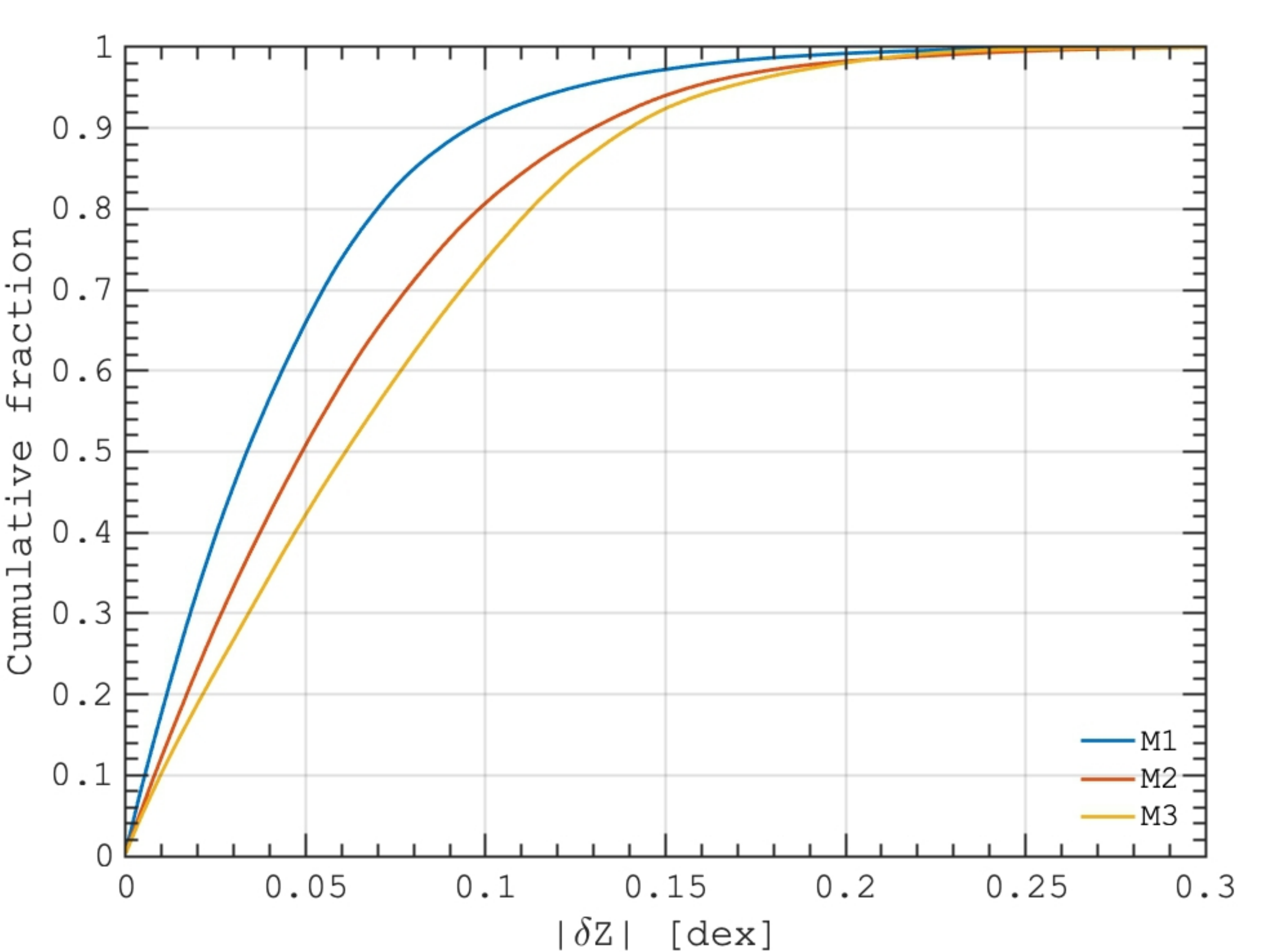}
\caption{Cumulative distribution of the residual metallicity~(after subtraction of the radial metallicity gradient) in different models.}\label{fig::delta_z_cumulative}
\end{center}
\end{figure}

\begin{figure*}
\begin{center}
\includegraphics[width=1\hsize]{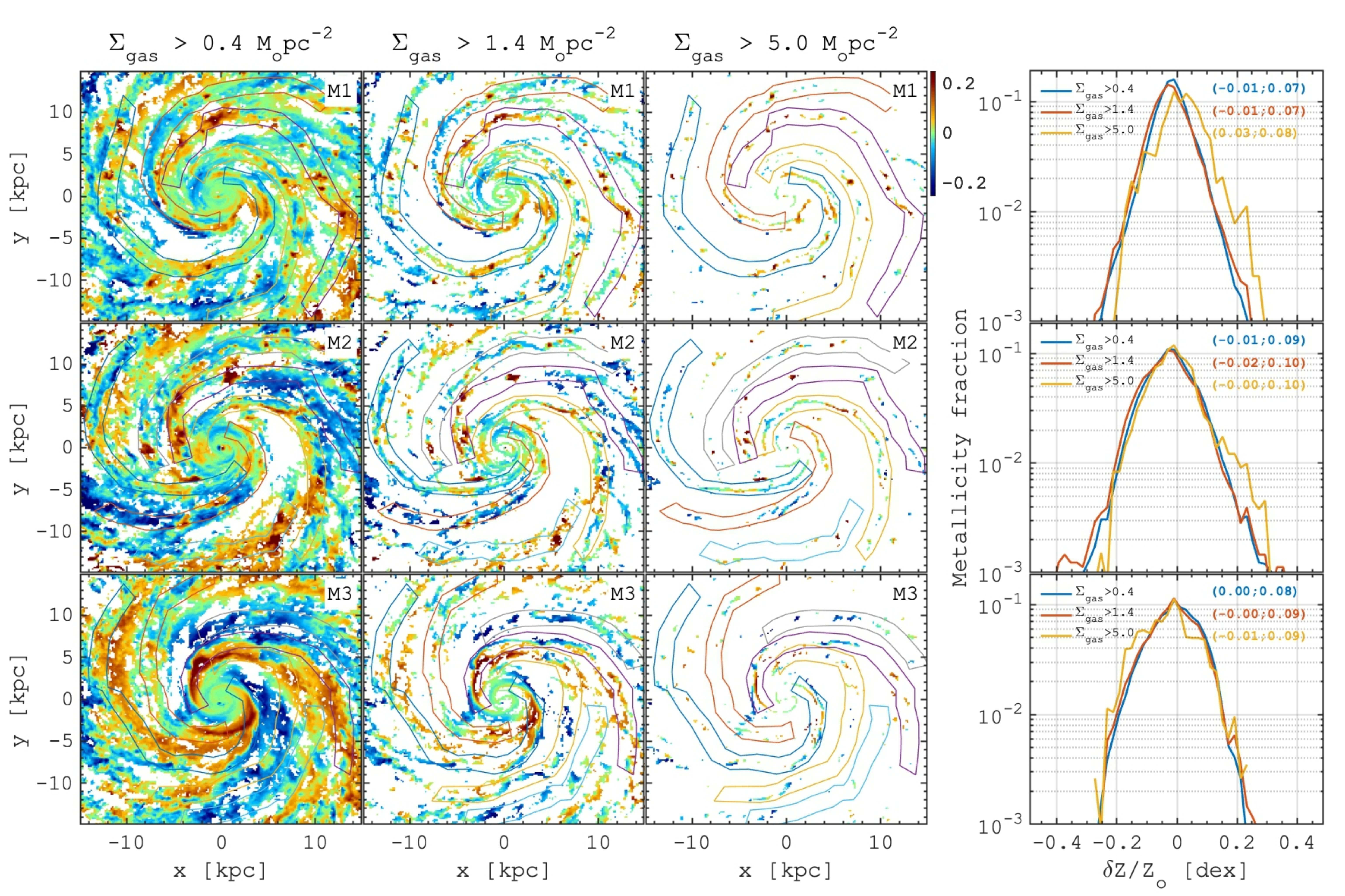}
\caption{The residual ISM metallicity maps where the regions with different minimal gas surface density are shown, from left to right: $\Sigma_{gas}>0.4$~\Msunpc, $\Sigma_{gas}>1.4$~\Msunpc and $\Sigma_{gas}>5$~\Msunpc. The rightmost column shows the distribution of the residual metallicity in the corresponding disk regions. Once the regions with lower gas density are masked the remaining ares follow the stellar spiral arms much close, however, the residual metallicity distributions are weakly affected by these selections. The pairs of numbers in the rightmost column show the mean and the standard deviation of the residual metallicity distributions.}\label{fig::metallicity_den_threshold}
\end{center}
\end{figure*}

\section{Models}\label{sec::models}

\subsection{Simulations setup}\label{sec::appendix_model}
We performed three $N$-body/hydrodynamical simulations of disk galaxies with a total stellar mass and a rotation curve compatible with those of the MW. All three models are the same in terms of the mass model but differ from each other by the initial radial profile of metallicity of the ISM and on/off enrichment of the ISM by newly formed stellar populations. Our models start from a pre-existing axisymmetric stellar disk with gas where we allow the star formation which is complemented by the chemical evolution of stellar populations.

In all three models, $5\times 10^6$ stellar particles are initially redistributed following a Miyamoto-Nagai density profile~\citep{1975PASJ...27..533M} that has a characteristic scale length of $3$~kpc, vertical thicknesses of $0.2$~kpc and mass of $6 \times 10^{10}$~\Msun. Our simulation also includes a live dark matter halo~($5\times 10^6$ particles) whose density distribution follows a Plummer sphere~\citep{1911MNRAS..71..460P}, with a total mass of $4\times 10^{11}$~\Msun ~and a radius of $21$~kpc. Gas component is represented by an exponential disk with a scale length of $5$~kpc and the total mass of $2\times 10^{9}$~\Msun. Gas dynamics is treated on a Cartesian grid with $5$~pc uniform spatial resolution. The initial equilibrium state has been generated using the iterative method from AGAMA software~\citep{2019MNRAS.482.1525V}.

In our simulations subgrid physics implementation is the same as in \cite{2021MNRAS.501.5176K}. In particular, we include the formation of new star particles which inherit both kinematics and elemental abundances of their parent gas cells. At each time step, for newly formed stars we calculate the amount of gas returned, the mass of the various species of metals, the number of SNII or SNIa for a given initial mass and metallicity, the cumulative yield of various chemical elements, the total metallicity, and the total gas released. Feedback associated with the evolution of massive stars is implemented as an injection of thermal energy in a nearby gas cell proportional to the number of SNII, SNI and AGB stars. The hydrodynamical part also includes gas-metallicity depended radiative cooling~\citep[see][for details]{2021MNRAS.501.5176K}.

Since our simulations aim to explore the formation of the azimuthal metallicity variations in three models we test the impact of the local ISM enrichment with and without pre-existing radial metallicity gradient in the gas. Models M1 and M2 include the ISM enrichment, while in model M3 we turn off the metals release to the ISM by newborn stellar populations. Models M2 and M3 start from the initial negative metallicity gradient~($\rm -0.15dex/r_e$, where $r_e=5.1$~kpc is the disk effective radius or $\rm -0.35dex/R_{25}$, where $R_{25}=12$~kpc) while model M1 has a constant initial metallicity of the gas. Therefore, Model M1 allows us to test how much the local enrichment alone is responsible for the azimuthal variations of the gas metallicity. Model M3 aims to quantify how the spiral arms induced redistribution of metals drives the azimuthal gradients, while Model M2 combines both effects. 

The simulations were evolved with the $N$-body+Total Variation Diminishing~(TVD) hydrodynamical code~\citep{2014JPhCS.510a2011K}. For the $N$-body system integration and gas self-gravity, we used our parallel version of the TREE-GRAPE code~\citep[][]{2005PASJ...57.1009F} with multithread usage under the SSE and AVX instructions. For the time integration, we used a leapfrog integrator with a fixed step size of $0.1$~Myr. In the simulation we adopted the standard opening angle $\theta = 0.7$. In recent years we already used and extensively tested our hardware-accelerator-based gravity calculation routine in several galaxy dynamics studies where we obtained accurate results with a good performance~\citep{2017MNRAS.468..920K,2018MNRAS.481.3534S,2020A&A...638A.144K}. 

\subsection{Pattern speed measurements}\label{sec::appendix1}
In this section, we provide details about the pattern speed measurements for individual spiral arms. In order to measure the pattern speed of the spiral arms, we analyse the stellar surface density morphology in three snapshots, one is the referenced one at $t_0$ and the other two correspond to $\rm t_0 \pm 1~Myr$. Thanks to the small interval between the snapshots we can directly measure how much different radial segments of the individual spirals rotated in azimuthal direction over $1$~Myr. In practice we split each individual spiral arm into $\delta \Rgal = 0.1$~kpc segments and calculate the following parameter:
\begin{equation}
\Oo \rm \epsilon(\phi) = \sum_{x,y} \left(\Sigma_0(x,y) - \Sigma^*(\phi,x,y)\right)^2\,,\label{eq:eps}
\end{equation}
where $\Sigma_0(x,y)$ is the stellar surface density at $t_0$ and $\Sigma^*(\phi,x,y)$ is the stellar surface density at $\rm t_0-1~Myr$~(or $\rm t_0+1~Myr$) rotated by $\phi$ in the azimuthal direction. Since the amplitude of the spiral arms does not change much over $1$~Myr, as the result of the procedure for each segment of the spiral arms we have a curve $\epsilon(\phi)$ with a global minimum at the angle corresponding to the best similarity between the arm at $t_0$ and $t_0\pm1$~Myr~(see Fig.~\ref{fig::xi}~(left) for a single spiral arm in M1 model). In other words, we are trying to find the angle of rotation of a given spiral arms segment over $1$~Myr~(backward and forward in time). Therefore, the angular offset corresponding to the minima of the curves in Fig.~\ref{fig::xi}~(left) divided by $1$~Myr results in the pattern speed of the spiral arms segments, or the pattern speed of individual arms as the function of \Rgal~(see the right panel). Since we have an opportunity to measure the rotation of spirals backwards and forward in time we present the curves $\epsilon(\phi)$ for $\rm t_0 - (t_0+1Myr)$ and $\rm (t_0+1Myr) - t_0$ where the minima for each \Rgal are marked by circles and squares, respectively.

In Fig.~\ref{fig::Omega0}~(right) we show the pattern speed of the individual arms~(crosses) compared to the rotational frequency~(circular velocity divided by \Rgal, solid lines) shown for each pattern speed by the same colour. For a better visibility, both the pattern speed and the corresponding rotational frequency are shifted by $10~\kmps$ for different arms. As we can see, the spiral arms in our model, similar to a number of other $N$-body/hydrodynamical simulations~\citep[see, e.g.,][]{2011ApJ...735....1W,2012MNRAS.426..167G,2013MNRAS.432.2878R}, do not rotate with the same pattern speed along the radius. This behaviour makes our results qualitatively similar to the ones presented in \cite{2016ApJ...830L..40S}.

\section{Results}\label{sec::results}

\subsection{Spiral structure properties and small-scale metallicity behaviour}

For each simulation, we focus our analyses on a single snapshot at $t_0=0.6$~Gyr of evolution, when a well-formed spiral structure is present. At later times the models are unstable to bar formation which substantially impacts both stars and the ISM, analyses of which is beyond the scope of the present study. In Fig.~\ref{fig::fig1} we show the face-on distributions of stellar~(first column) and gas density~(second column) together with the mean ISM metallicity~(third column), the residual ISM metallicity~(the fourth column, after the subtraction of the radial metallicity gradient (i.e. the mean metallicity at a given \Rgal shown by the white solid lines in the fifth column) and the radial metallicity profile~(fifth column). In the fifth column we also provide the exponential fits of the radial metallicity profiles~(red solid lines, shifted vertically). The slopes of the ISM metallicity we measure, while being mainly inherited from the initial setup, are comparable to the ones observed in the nearby disk galaxies~\citep{2014A&A...563A..49S, Belfiore2017,2019ApJ...887...80K,Zinchenko2019,Zinchenko2021,Zurita2021}.

Although our initial setup~(mass model and initial equilibrium state) is the same in all the models, the morphology of the spiral structure is slightly different at the same snapshot in time. This is likely the result of stochasticity of the growth of the perturbations in stellar-gaseous disks~\citep[see, e.g.,][]{2011MNRAS.416.1191R} due to different ISM dynamics caused by metallicity-dependent gas cooling rates. Nevertheless, a Fourier analysis of the snapshots suggests that the modal composition of the spiral structure is very similar~(see the radial structure of Fourier harmonics in Fig.~\ref{fig::fourier_amps}~(top row)) and all three models reveal a multi-arm, tightly wound spiral structure similar to the one usually obtained in $N$-body simulations. It is seen that the spiral arms structure, revealed by the Fourier analysis, is dominated by the $m=2$ and $m=3$ modes, which superposition results in a slightly different morphology among our models~(see Fig.~\ref{fig::fig1}~(left)) at a given time.

In order to quantify the ISM metallicity behaviour in the vicinity of the spirals we localize the individual arms which we define as the positive part of 2D stellar density perturbation~(overdensity): 
\begin{equation}
\Oo \rm \delta \Sigma = \frac{\Sigma(R_{gal},\varphi) - \left\langle\Sigma(R_{gal},\varphi )\right\rangle_\varphi}{\left\langle\Sigma(R_{gal},\varphi)\right\rangle_\varphi}\,,\label{eq::eq1}
\end{equation}
where \Rgal and $\varphi$ are the radius and azimuth in cylindrical coordinates and the brackets $\left\langle\right\rangle_{\varphi}$ indicate the mean -- i.e. azimuthally averaged value at a given \Rgal. In Fig.~\ref{fig::fig1} the stellar spiral arms are highlighted by coloured lines spanning $2$~kpc radial \Rgal range. Although, the ISM morphology is quite complicated, because of a number of chains of giant clumps near the arms and isolated clouds in between the arms~\citep[see, e.g.,][]{2006MNRAS.371.1663D,2013MNRAS.436.1836R,2014MNRAS.439..936F,2016MNRAS.455.1782K,2017MNRAS.468..920K}, one can see that inside $\Rgal\lesssim10$~kpc the maxima of the large-scale gas density distribution correspond to the leading side of the stellar spiral arms, as it has been expected for slowly rotating spirals~\citep[see, e.g.,][]{2006MNRAS.367..873D,2011AstL...37..563K,2016MNRAS.458.3990P}. 

In order to test the impact of the local enrichment and the radial gradient transformation on the azimuthal variations of metallicity, one needs to be sure that both the strength and pattern speed of the spiral arms are the same in different models. We showed that the amplitudes of the stellar density perturbations are essentially the same in all the models~(see Fig.~\ref{fig::fourier_amps}, top). Also, the spiral arms for all three models rotate slower than the gas in the inner disk~($\lesssim8-10$~kpc) and corotate with the gas in the outer disk~(see Fig.~\ref{fig::Omega0}, the right panel). Therefore, the material stellar arms are non-steady; they are wound and, likely, stretched by the galactic shear in the outer disk. In the inner disk, spirals can bifurcate and merge with the others~\citep[see, e.g.,][]{2011MNRAS.410.1637S,2011ApJ...730..109F}. This makes the ISM structures and abundance patterns associated with the stellar arms also non-steady.

\begin{figure}[t]
\begin{center}
\includegraphics[width=1\hsize]{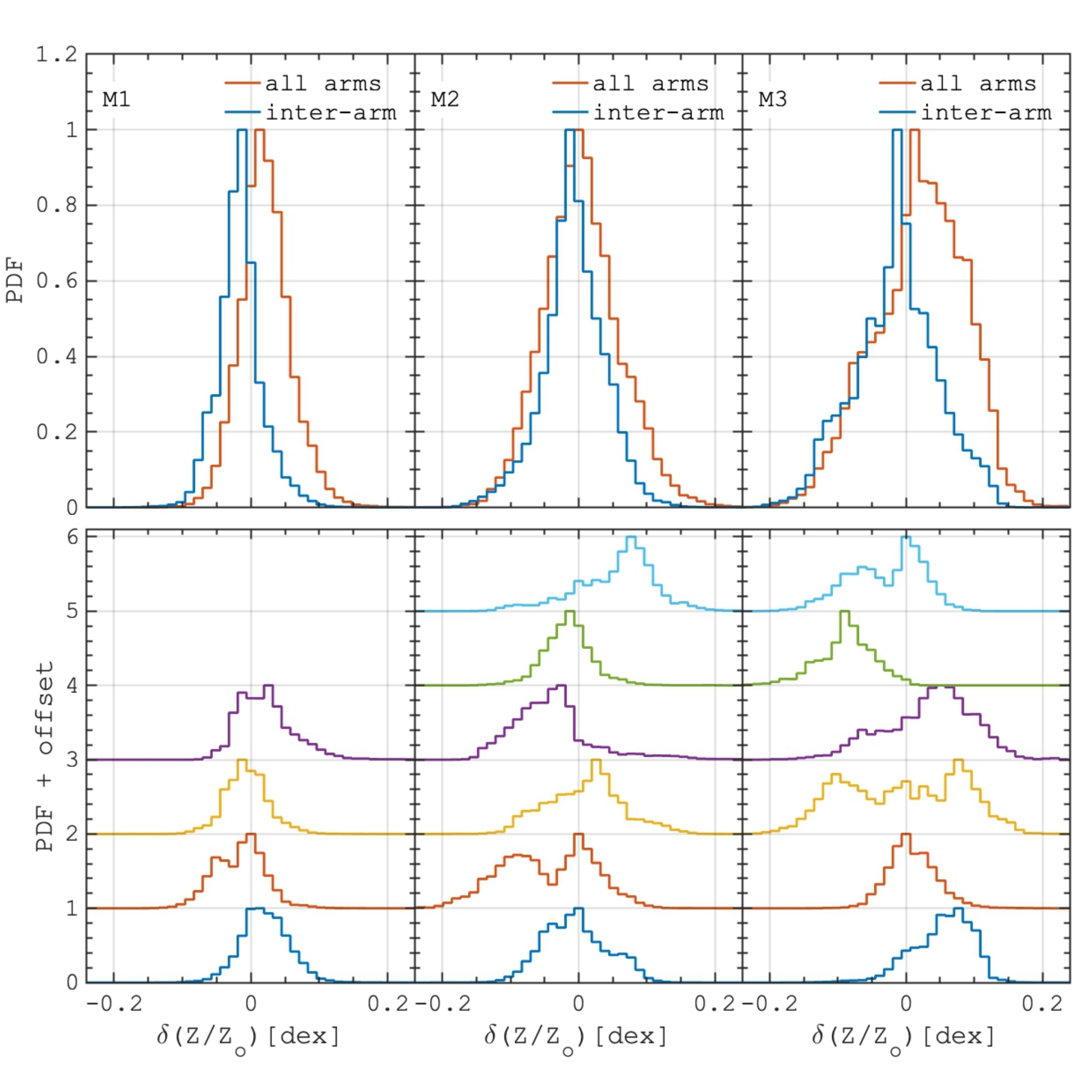}
\caption{Probability distribution functions for the residual ISM metallicity in arms and inter-arm regions~(top). Bottom row shows the corresponded PDFs for individual spiral arms where the colour of the lines is the same as in Fig.~\ref{fig::fig1}.}\label{fig::fig2}
\end{center}
\end{figure}

\begin{figure}[t]
\begin{center}
\includegraphics[width=1\hsize]{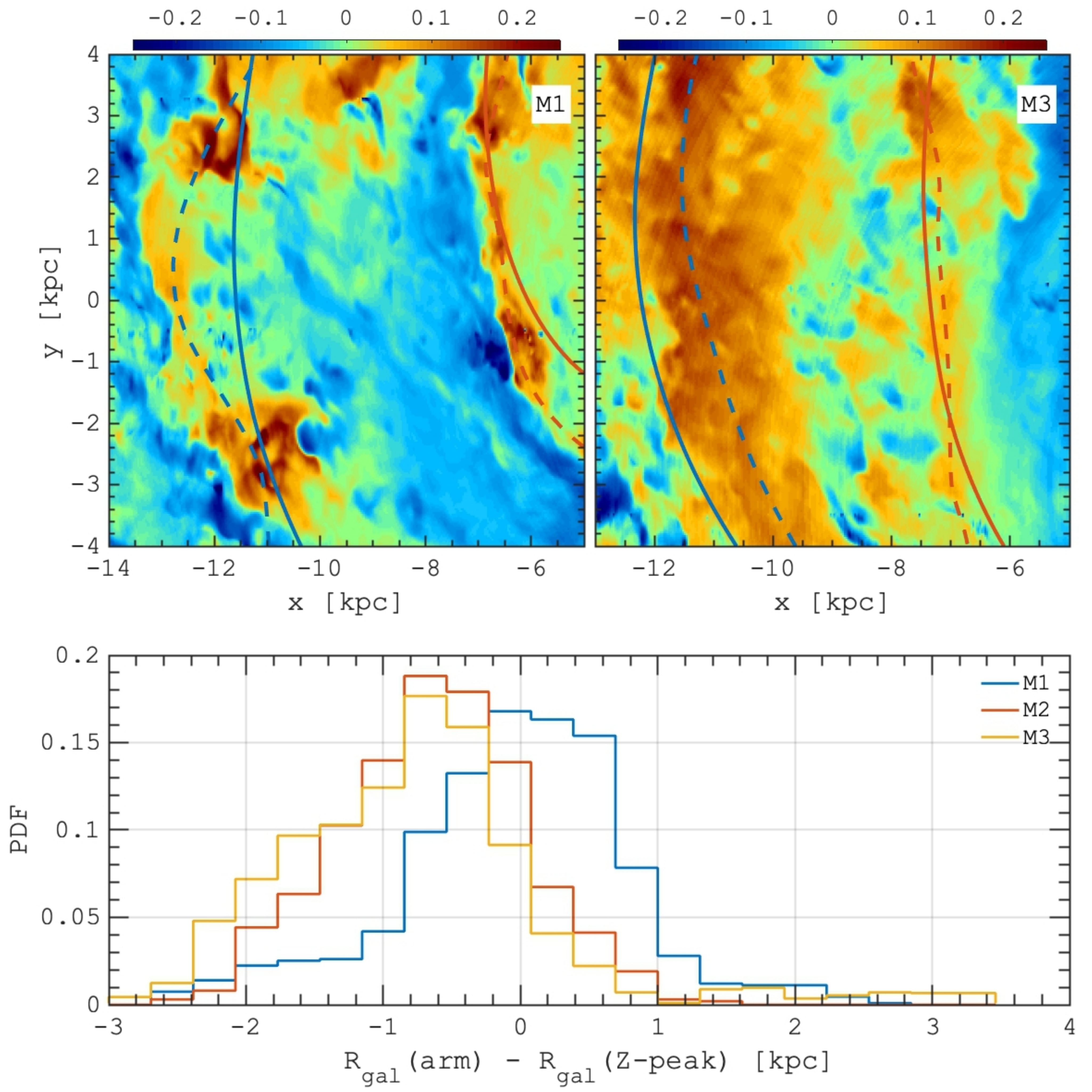}
\caption{{\it Top:} background maps are the residual ISM metallicity in model M1~(no radial gradient, left) and M3~(no ISM enrichment, right) where the location of the stellar spiral arms are shown by solid lines while the mean location of the metallicity peaks -- by dashed lines. {\it Bottom:} distribution of the radial offset between the mean location of the stellar spiral arms and the metallicity peaks for different models.}\label{fig::fig4}
\end{center}
\end{figure}

In Fig.~\ref{fig::fig1}~(see the third column) the mean metallicity distributions show a spiral-like morphology which however differs from the ones seen in gas and stars. To quantify the strength of the azimuthal metallicty variations, similar to the stellar overdensity~(see Eq.~\ref{eq::eq1}) we derive the residual metallicity:
\begin{equation}
\Oo \rm \delta Z/Z_\odot = \frac{ (Z/Z_\odot)(R_{gal},\varphi) - \left\langle(Z/Z_\odot)(R_{gal},\varphi)\right\rangle_\varphi }{ \left\langle(Z/Z_\odot)(R_{gal},\varphi)\right\rangle_\varphi}\,,\label{eq::eq2}
\end{equation}
where $\left\langle(Z/Z_\odot)(R_{gal},\varphi)\right\rangle_{\varphi}$ is the azimuthally-averaged radial metallicity profile~(see the white lines in Fig~\ref{fig::fig1}, fifth column). Maps in Fig.~\ref{fig::fig1}~(fourth column) demonstrate that the residual metallicity varies from negative to positive values across the arms. In particular, we find that at a given \Rgal the scatter is about $0.05$~dex which is similar to the numbers obtained in different observations~\citep[e.g.,][]{Zinchenko2016,2019ApJ...887...80K}. To demonstrate that in Fig.~\ref{fig::delta_z_cumulative} we show the cumulative distribution of the absolute values of the residual metallicity for different models. The figure shows that the metallicity deviates from the mean value at a given radius by $0.05$~dex for $40-65$\% of cases.

\subsection{Large-scale metallicity behaviour across the spiral arms}

In Fig.~\ref{fig::fig1} we notice that in M1 the residual metallicity does not reveal a large-scale spiral pattern, but instead consists of a number of patchy segments, while in models M2 and M3 the regions of systematically lower~(higher) metallicity span across the entire disk. This is highlighted more quantitatively in Fig.~\ref{fig::fourier_amps} where the radial profile of Fourier coefficients are presented. In these models one can see a dominant $m=2,3$ mode~(more prominent in negative $\delta Z$, fourth column in Fig.~\ref{fig::fig1}) despite the presence of six spiral arms. This result suggests that the large-scale metallicity pattern, in the presence of radial gradient, likely follows the dominant spiral mode~(see Figs.~\ref{fig::fig1},~\ref{fig::fourier_amps}).

In observations, the information about the ISM metallicity distribution is often based on the data from the \hii regions~\citep[see, e.g.,][]{1996ApJ...456..504K,2012A&A...545A..43C,2017A&A...601A..61V,2019ApJ...887...80K}, where the gas density is high enough to sustain recent star-formation. This usually does not allow for a mapping of the metallicity distribution everywhere in the disk and limits the analysis to some sparsely distributed regions. In our simulations we are not able to resolve individual \hii regions, however, to link the observational results with our models we first limit the analysis of the ISM metallicity to the regions with high gas density. In Fig.~\ref{fig::metallicity_den_threshold} we show the distribution of the residual metallicity where we masked the regions with the gas surface density below given values~($0.4$, $1.4$ and $5$~\Msunpc). Obviously, once we move to the metallicity distribution in regions with higher gas density the coverage of the disk decreases and the remaining regions trace better the spiral arms. However, there is no prominent systematics in the residual metallicity values in the remaining high-gas density regions. In particular, the second and third columns of Fig.~\ref{fig::metallicity_den_threshold} show that the residual metallicity varies in a wide range where the entire arms or their small patches can have systematically either positive or negative values of the residual metallicity. If, following some theoretical expectations~\citep{2016ApJ...830L..40S,2017ApJ...846...39H,2019A&A...628A..38S}, the high-metallicity ISM is associated with the spiral arms then the mean of the metallicity distribution should move towards positive values once we mask low-gas density regions. To test this in the rightmost column in Fig.~\ref{fig::metallicity_den_threshold} we show the distribution of the residual metallicity in the regions with high gas density. In other words, we plot the distribution of the values from maps shown in the first three columns of Fig.~\ref{fig::metallicity_den_threshold}. The resulting distributions however do not vary significantly, and both the mean and its dispersion are weakly impacted by the spatial selections based on the gas surface density. This suggests that similar to a number of observational studies, there is no apparent match of the large-scale metallicity patterns with the spiral arms in our models. However, since Fig.~\ref{fig::fig1} shows the metallicity distributions still demonstrating certain patterns, in the following, we analyse the complete metallicity distributions without masking any low gas-density regions.

To analyze the variations of the metallicity more quantitatively, in Fig.~\ref{fig::fig2} we show the residual metallicity distributions for all spiral arms and inter-arm region~(top) and for individual spiral arms~(bottom). We see that the residual metallicities for both arm and inter-arms regions vary in roughly the same range, while spiral arm regions have slightly higher metallicities. The effect is the most prominent in Model M3~(without enrichment) suggesting that the release of metals by newly formed stars does not correlate much with the spiral arms but tends to smear the metallicity distribution at a given \Rgal~\citep[see, e.g.,][]{2012ApJ...758...48Y,2018MNRAS.475.2236K}.

The distributions in Fig.~\ref{fig::fig2} suggest that systematically negative residuals in metallicity do occur in some of the spiral arms. This contradicts naive expectations that star formation would locally increase the ISM metallicity. To search for the latter effect, we find the location of nearby positive peaks in metallicity next to the arms and, thus, measure the offset to the spiral arms. In Fig.~\ref{fig::fig4}~(top) we show two examples of the offset calculation for the pair of arms in M1 and M3 models. One can see that there is a positive~(larger \Rgal) offset of the metallicity peak~(dashed lines) to the blue arm~(solid) in M1 while in model M3 we see the opposite configuration~(peak of metallicity has systematically smaller \Rgal relative to the stellar spiral arm). For the red arms in both models, the offset is, on average, zero. Of course, the offset we calculate may not represent a generic connection between the spiral arms and metallicity distribution, especially at local scales, however, it gives an idea of how, on average, the spatial distribution of metals correlates with the spiral arms depending on initial radial gradient and ISM enrichment. In the bottom panel of Fig.~\ref{fig::fig4}, we show the distribution of the offset for all the arms in three models. Models with initial radial gradient~(M2, M3) show on average a negative radial offset between spiral arm and metallicity peak, suggesting the presence of higher metallicity behind the trailing arms. In M1, the offset distribution is roughly symmetric. This behaviour is likely linked to the SF activity where newborn stars can release metals far away from their birthplaces in the spirals or even reach other arms, especially in the inner disk, where spirals rotate slower than the gas, thus breaking a coherent cycle of enrichment/mixing proposed in some previous works~\citep{2017ApJ...846...39H,2019A&A...628A..38S}. 

\begin{figure}[t]
\begin{center}
\includegraphics[width=1\hsize]{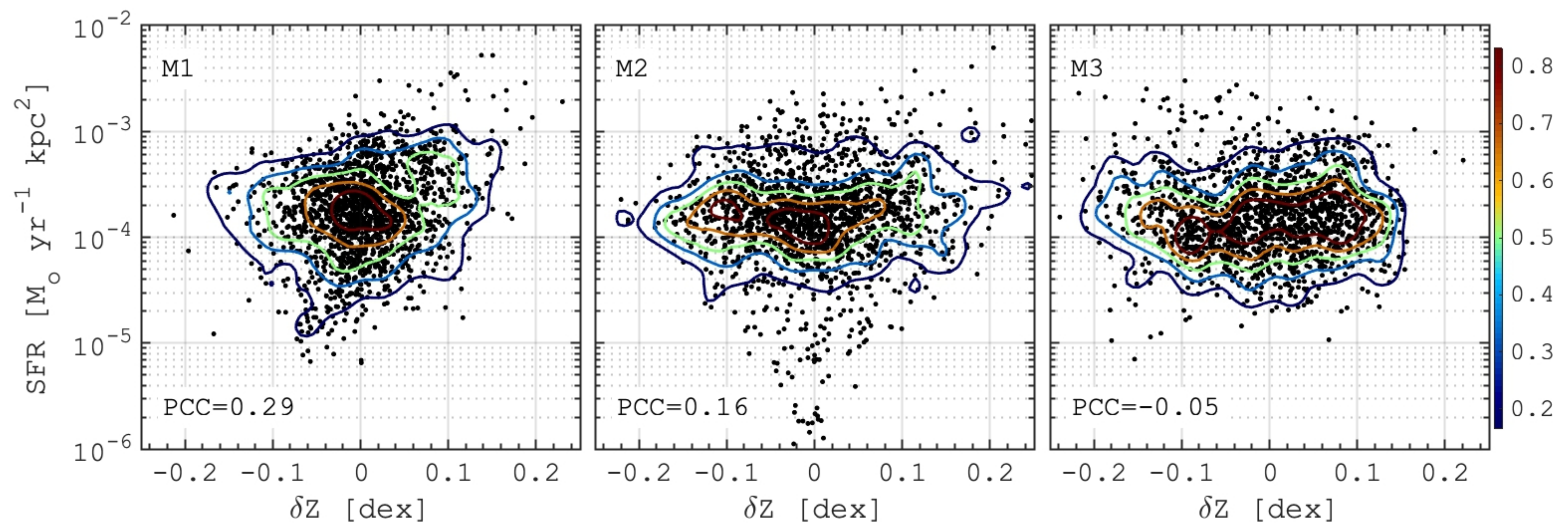}
\caption{Relations between the residual metallicity and star formation surface density. Coloured lines represent the density contours. The corresponding Pearson correlation coefficients are given in each panel.}\label{fig::sfr_z}
\end{center}
\end{figure}

To test further the hypothesis about the impact of the local enrichment, in Fig.~\ref{fig::sfr_z} we show the relation between the SF surface density and the metallicity residuals. The relation is based on the $\rm 500\times500~pc^2$ smoothing of both SFR and $\delta Z$ metallicity XY-maps.  Although we do not see a strong correlation for any of our models, in M1, there is a clear trend where more metal-rich regions tend to be spatially associated with more intense SF. Finally, we test how much the ISM dynamics affects the appearance of the large-scale abundance variations in the azimuthal direction. 

\begin{figure}[t]
\begin{center}
\includegraphics[width=1\hsize]{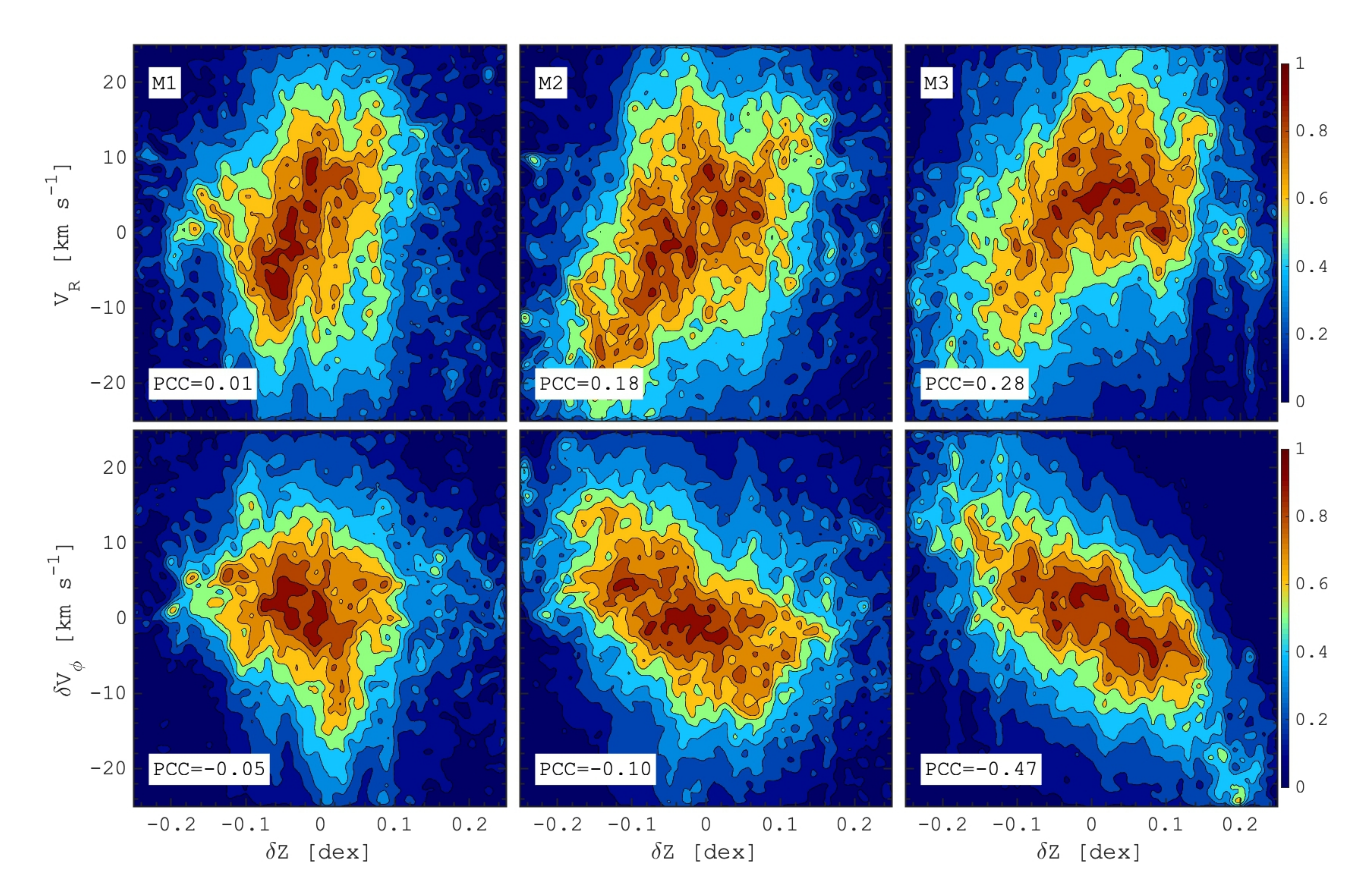}
\caption{Relations between the residual metallicity and gas radial velocity component~(top) and residual azimuthal velocity component~(subtracted circular velocity, bottom). The colours represents the mass fraction of the gas.}\label{fig::vrvphi_z}
\end{center}
\end{figure}

In Fig.~\ref{fig::vrvphi_z} we compare the relation between both  radial~($V_R$) and residual azimuthal velocity~($\delta V_\varphi$) components of the gas with the residual metallicity. The residual azimuthal velocity component is the gas velocity component after the subtraction of the mean azimuthally-averaged rotation~(rotation curve, see Eq.~\ref{eq::eq1} where $\Sigma$ should be replaced by $V_\varphi$). We see no correlation in the case of M1~(only enrichment) and a very clear correlation for M3~(no enrichment, but initial radial gradient)  while model M2 shows an intermediate behaviour. Similar to the $\rm SFR-\delta Z$ analysis in Fig.~\ref{fig::sfr_z}, we calculated the Pearson correlation coefficients which allow us to quantify the relations between the gas velocity components and the residual ISM metallicity~(see numbers in the panels of Fig.~\ref{fig::vrvphi_z}). This result suggests that, even slowly~(compared to the gas) rotating spiral arms drag a more metal-rich gas from the inner parts of the disk to the trailing side of the arms and, thus, increasing the mean metallicity of the ISM behind the spirals. This picture explains why in models M2 and M3 we find a prominent negative offset of the metallicity peaks relative to the spiral arms~(see Fig.~\ref{fig::fig4}).  

\vspace{-\topsep}\section{Summary}\label{sec::conclusions}

Using hydrodynamical simulations of isolated spiral galaxies, we study the impact of the local enrichment and pre-existing radial metallicity gradient transformation on the formation of azimuthal metallicity variations in the vicinity of the spiral arms. Similar to some previous studies, in our models, the pattern speed of individual arms varies with the radius being slow~(compared to the disk rotation) in the inner~($<8$~kpc) and corotating in the outer disk. Our main results are as follows.

\begin{itemize}

\item[-] Both types of models with local enrichment and pre-existing radial abundance gradient are able to produce the azimuthal scatter of the ISM metallicity~($\approx0.05$~dex) in simulated spiral galaxies~(see Fig.~\ref{fig::fig1}). Analysis of the ISM metallicity as a function of the underlying gas density does not reveal any strong relations between the residual metallicity and the location of the stellar spiral arms~(see Fig.~\ref{fig::metallicity_den_threshold}). Although individual arms could have systematically lower or higher~(than the mean) metallicity, gas in the spiral is slightly more metal-rich compared to the inter-arms region~(see Fig.~\ref{fig::fig2}).

\item[-]  We find that model with local enrichment only~(M1, no pre-existing metallicity gradient) is not able to reproduce large-scale spiral arms-like variations of the mean metallicity~(see Figs.~\ref{fig::fig1},~\ref{fig::fourier_amps}). In this model, the short-scale patches of high-metallicity regions can be found suggesting that the enrichment of the ISM does not correlate much with the large-scale spiral structure.

\item[-] Both models with pre-existing radial metallicity gradient~(M2 and M3) show the formation of the spiral-like metallicity pattern but its morphology differs from the stellar spiral structure~(see Figs.~\ref{fig::fig1}). Although we identify a six-armed spiral structure, the large-scale residual metallicity pattern depicts $m=2,3$ structure corresponding to the most significant 2D Fourier harmonics of the stellar density distribution~(see Fig.~\ref{fig::fourier_amps}).

\item[-] We found a substantial radial offset between spiral arms and the metal-rich ISM pattern. The amplitude of the offset reaches up to $1-1.5$~kpc~(see Fig~\ref{fig::fig4}). In models with the radial gradient we find no correlation of the residual metallicity with recent star-formation while in model M1~(without) gradient we can see a weak increase of the metallicity with higher star formation.
Model without ISM enrichment shows the most prominent correlation of the residual metallicity with the gas velocity components: larger metallicity corresponds to larger negative~(inflow) radial velocities and negative azimuthal velocity residuals. Therefore, we suggest that dynamical effects play a key role in the formation of the large-scale metallicity variations across spiral arms.

\end{itemize} 

Our models, while being rather simplified, nevertheless allow us to propose an explanation of the observational data demonstrating a controversial picture of the systematic azimuthal variations of metallicity around spiral arms. We suggest that the ISM enrichment near the arms alone is unlikely responsible for the systematic azimuthal metallicity pattern, at least in the case of non-steady spirals, while the key ingredient is a pre-existing radial abundance gradient. If the observed radial gradients are small~\citep{2014A&A...563A..49S}, they are likely not enough to be transformed into the azimuthal variations in most of the galaxies. However, once the azimuthal variations are found, more likely they will depict the shape of the most significant spiral mode~($m=2, 3$), which can explain prominent oxygen variations found in some barred galaxies~\citep{2016ApJ...830L..40S,2017ApJ...846...39H}, however, in some cases flattening of the radial gradient should act against the formation of the systematic azimuthal variations. 

Extending the model presented in \cite{2016ApJ...830L..40S} we suggest that not only corotating spirals are responsible for the azimuthal variations of the ISM metallicity but also slowly~(compared to the gas) rotating patterns provide similar results. Obviously, our predictions will depend on the strength and nature of the spirals and also star formation activity, which, along with the Fourier analysis of the 2D metallicity distribution, may be tested by IFU surveys in the near future.

\begin{acknowledgements}
SK deeply appreciates Igor Zinchenko (LMU, Munich; MAO, Kyiv) for his contribution to the early versions of the paper. The authors thank the anonymous referee for a constructive report. SK also thanks Daisuke Kawata for useful discussion about the nature of corotating spirals. Numerical simulations were carried by using the equipment of the shared research facilities of HPC computing resources at Lomonosov Moscow State University~(project RFMEFI62117X0011) supported by the Russian Science Foundation~(project no. 19-72-20089). 
\end{acknowledgements}

\bibliographystyle{aa}
\bibliography{GasMetallicity}

\end{document}